\documentclass[twocolumn, linenumbers]{aastex631}
\nolinenumbers 
\usepackage{gensymb}

\begin{document}

\title{Spin-down changes in PSR B0540-69 induced by a drift of the magnetic axis}

\author[0009-0000-3268-9058]{Lucas G. Barão}
\author[0000-0003-4089-3440]{J. E. Horvath}

\affiliation{Universidade de São Paulo (USP), Instituto de Astronomia, Geofísica e Ciências Atmosféricas (IAG)\\
R. do Matão, 1226, Cidade Universitária\\
São Paulo, SP 05508-090, Brazil}

\begin{abstract}
\nolinenumbers
The dynamics of the solid crust + magnetic field lines of pulsars is a much debated issue, and remains unsettled after 50 years. Some pieces of evidence have emerged to complete and confirm theoretical calculations and expectations. We discuss in the present work an interpretation of the behavior of the ``Crab Twin'' pulsar PSR B0540-69 in terms of the evolution of the magnetic field/quakes, connecting the behavior of the braking index with the underlying platelet drift and sudden discontinuous rearrangement (fast-slip) and long-term ones (slow-slip events), suggested by analogy with existing theoretical picture observed in the Earth's crust. The relationship of this scenario with permanent torque-changing glitches seen in the Crab and other young pulsars, and a set of similar events in the same object and others is addressed. We conclude that this physical approach is in principle consistent with all these sudden events, and point out future work to clarify the whole picture.
\end{abstract}


\section{Introduction}
\label{sec:intro}
Discovered in 1967, pulsars were eventually identified with rotating neutron stars in which a large magnetic field establishes a 
magnetosphere of accelerated co-rotating charges near the surface. Each time a revolution is completed, a pulse is produced in a yet 
unconfirmed manner (see \citet{philippov2022pulsar} for a review of pulsar emission mechanisms). In addition, several pulsars were later identified in optical, X-rays and gamma rays, and theories of these higher energy emission developed \citep{petri2016theory}. The whole picture of these emissions is needed to understand the pulsar phenomenon. 

One of the first observables identified as an useful tool to check the pulsar dynamics is the so-called {\it braking index}, namely 
the combination of the frequency and its time derivatives constructed as

\begin{equation}
n = \frac{\Omega {\ddot \Omega}}{{\dot \Omega}^2}.
\label{eq:braking}
\end{equation}

If we assume the standard idea that the only mechanism causing a pulsar to lose energy is the magnetic dipolar radiation, and all the structural parameters are constant (i.e., not varying with time), then the dynamical equation obtained by the integration of the flux of the Poynting vector crossing a sphere centered at the object yields

\begin{equation}
    I\dot\Omega=-\frac{2B^2R^6\sin^2\chi}{3c^3}\Omega^3
    \label{eq:magneticdipole}
\end{equation}

with $I$ the moment of inertia of the neutron star, $R$ its radius, $B=|\vec{B}|$ the strength of the poloidal magnetic field on the surface of the star and $\Omega$ the angular velocity. The magnetic field and the rotation axis are assumed to form an angle $\chi = constant$ in this simple model. In this case, the braking index of Equation \ref{eq:braking} should be exactly 3. None of the empirically determined braking indexes complies with this expectation \citep{livingstone2011post,hamil2015braking}, and this indicates an incomplete knowledge of pulsar torques. This is why several observations and theoretical developments over the years have argued that this initial model is too simple to account for the complex phenomenology present in many objects. Episodes of enhanced braking have been observed and interpreted as an extra torque due to a temporary reconfiguration of the magnetosphere \citep{kramer2006periodically}. Other neutron stars supposedly having larger magnetic fields than ordinary pulsars (magnetars) have shown large flares suggesting the transient presence of a wind, a component not accounted for in Equation \ref{eq:magneticdipole}. In fact, the observation of a set of Pulsar Wind Nebula (PWN) has been also interpreted as stemming from a particle wind, although this component is rarely considered in the torque equation. 

Other timing irregularities have been related to {\it time-varying} quantities in the torque equation. For instance, small quasi-permanent changes in the braking of the Crab \citep{link1997we,allen1997glitches} and other pulsars after a glitch \citep{horvath2019braking, hu2023probing} have been suggested to arise from sudden ``jumps'' in the magnetic field $B$, the moment of inertia $I$ or the angle $\chi$ in Eq. \ref{eq:magneticdipole}. All three possibilities rely on physical arguments about the secular evolution of the pulsar/magnetosphere system. We shall focus in this work on the last one, related to a predicted drift of the magnetic axis considered in several works.

The fundamentals of the neutron star tectonics scenario have been elaborated long ago. In a work related to the previous idea of starquake-induced glitches \citep{1969}, \citet{Ruderman1991a} showed that in spinning-down fast pulsars, the solid crustal platelets, consistent of heavy nuclei and superfluid neutrons, can release a large amount of elastic energy when exceeding the maximum shear stress $\theta_{\rm{max}}$ they can support, possibly leading to GRBs, FRBs and glitch-like events \citep{OthersI, OthersII, Horvathetal2022}. A tangential crust displacement across the surface of the neutron star, towards the equator, carrying with it the local magnetic field would be present (since the scale of the plates is small compared to the star radius, we refer them as \textit{platelets} in this work). This rearrangement of the platelets would imply an increase in the inclination angle and a progressive conversion of the pulsar into an orthogonal rotator. It has even been theorized that this drift causes variations in the braking index, depending on the configuration of the geometry of the magnetic poles (see Fig. 5 in \citet{Ruderman1991b}). More recently, \citet{Gourgouliatosetal2016} showed that in an energy equipartition scenario between poloidal and toroidal magnetic fields in the neutron star, magnetic instabilities can transfer energy to non-axisymmetric, sub-$km$-sized zones, in which local field strength can greatly exceed that of the global-scale field. Such intense small-scale magnetic zones were shown to induce at some point high-energy bursts through local crust yielding. We interpret this mechanism as the analogous cause for the fast-slip events of the Earth's crust, which involve a number of platelets suddenly releasing a large amount of energy, leading in the case of neutron stars to magnetic field direction displacements of $10^{-5}-10^{-4}\;\mathrm{rad/yr}$ \citep{lander2019magnetic,gourgouliatos2022magnetic} after the event. In this way, we envisage the behavior of PSR B0540-69 as stemming from a plate dynamics analogous to the Earth's crust. This phenomenon can induce a variation within same order of magnitude as the alternative models (for a review of different inclination angle evolution, see \citet{LiGao}). In summary, plate dynamics is arguably the cause of a drift in the magnetic axis, and we shall elaborate and apply these ideas to the behavior of PSR B0540-69 below. \\

\begin{table*}[t]
 \centering
 \begin{tabular}{c|cccc}
        Date & $(T-T_0)/365\;\rm{(yrs)}$ & $n$ & $\dot{\nu} \;(10^{-10} \;\mathrm{Hz \,s ^{-1}})$ & $\ddot{\nu} \;(10^{-23} \;\mathrm{Hz \,s ^{-2}})$ \\ \hline
        Before SRT11 & - & $2.129\pm0.012$ & $-1.86$ & - \\
        12-xx-2011 & 0 & $0.031\pm0.013$ & $-2.53$ & 9.9 \\
        02-17-2015 & 3.85 & $0.11\pm0.09$ & $-2.53$ & 35.7\\
        11-05-2015 & 4.52 & $0.09\pm0.2$ & $-2.53$ & 29.2\\
        06-07-2016 & 5.06 & $0.2\pm0.1$ & $-2.53$ & 65.0\\
        01-02-2017 & 5.61 & $0.1\pm0.2$ & $-2.53$ & 32.5\\
        07-16-2017 & 6.11 & $0.5\pm0.3$ & $-2.53$ & 162.5\\
        02-11-2018 & 6.91 & $0.8\pm0.08$ & $-2.53$ & 260.0\\
        03-20-2019 & 7.81 & $1.2\pm0.2$ & $-2.53$ & 390.0\\
        04-16-2023 & 11.62 & $1.4$ & $-2.53$ & 467.0
 \end{tabular}
 \caption{PSR B0540-69 main properties changes after the spin-down rate transition. This data is collected from \citet{wang2020braking} and \citet{rusul2023external}. The second column is a time normalization starting from the SRT time, happened sometime in 2011 December 3rd and 17th. In the last detection, there was a change in $\dot\nu$ of the order of magnitude of $10^{-5}$.}
 \label{tab:brakingindex}
\end{table*}

\section{The behavior of PSR B0540-69}

The young pulsar PSR B0540-69 (a.k.a. ``the Crab Twin'') is one of the most intriguing and active objects known in the pulsar sample. It has been seen to undergo a large spin-down rate transition (SRT) event in 2011 (we shall write ``SRT11'' to refer to this event below) at $\sim$ constant frequency, but with changes in ${\dot \nu}$ and its second derivative ${\ddot \nu}$, which resulted in an abrupt change in the braking index, adopting a very different figure before and after the event \citep{wang2020braking}. Just like other young neutron stars, PSR B0540-69 displays a Pulsar Wind Nebula (PWN), revealing that energetic particles are being ejected from the magnetosphere, a phenomenon well-known from Crab and Vela, but which is generally dismissed in the energy emission analyses and secular dynamical evolution. 

The fundamental torque equation for this pulsar, in which a term representing the wind emission of relativistic particles from the star's surface \citep{HCK, LiGao, Sun, horvath2025evolution} has been introduced, is

\begin{equation}
    I\dot\Omega=-\frac{2B^2R^6\sin^2\chi}{3c^3}\Omega^3-BR^3\sin^2\chi\sqrt\frac{L_p}{2c^3}\Omega,
    \label{eq:torque}
\end{equation}

where $L_p$ is the \textit{particle luminosity}, the amount of energy (per unit of time) carried by relativistic particles in the magnetosphere. In a previous work \citep{horvath2025evolution} we applied this wind model to the newly discovered Long Period Sources \citep{hurley2022radio,hu2023probing,dong2024discovery} and shown that an additional amount of energy been dissipated along their life and can help explain these slow pulsars. Even though the wind term is $\propto \, \Omega$, when compared with the standard dipolar term $\propto \, \Omega^3$, the former spin-down mechanism is quite relevant for the energy loss in young pulsars, and the Crab Twin PWN is evidence that a significant amount of energy is lost not only by electromagnetic radiation emissions, but rather by relativistic particles escaping through open magnetic lines as well \citep{contopoulos2014new,petri2016theory}. Therefore, we proceed to analyze the SRT considering this component.

\begin{figure}
    \centering
    \includegraphics[width=\linewidth]{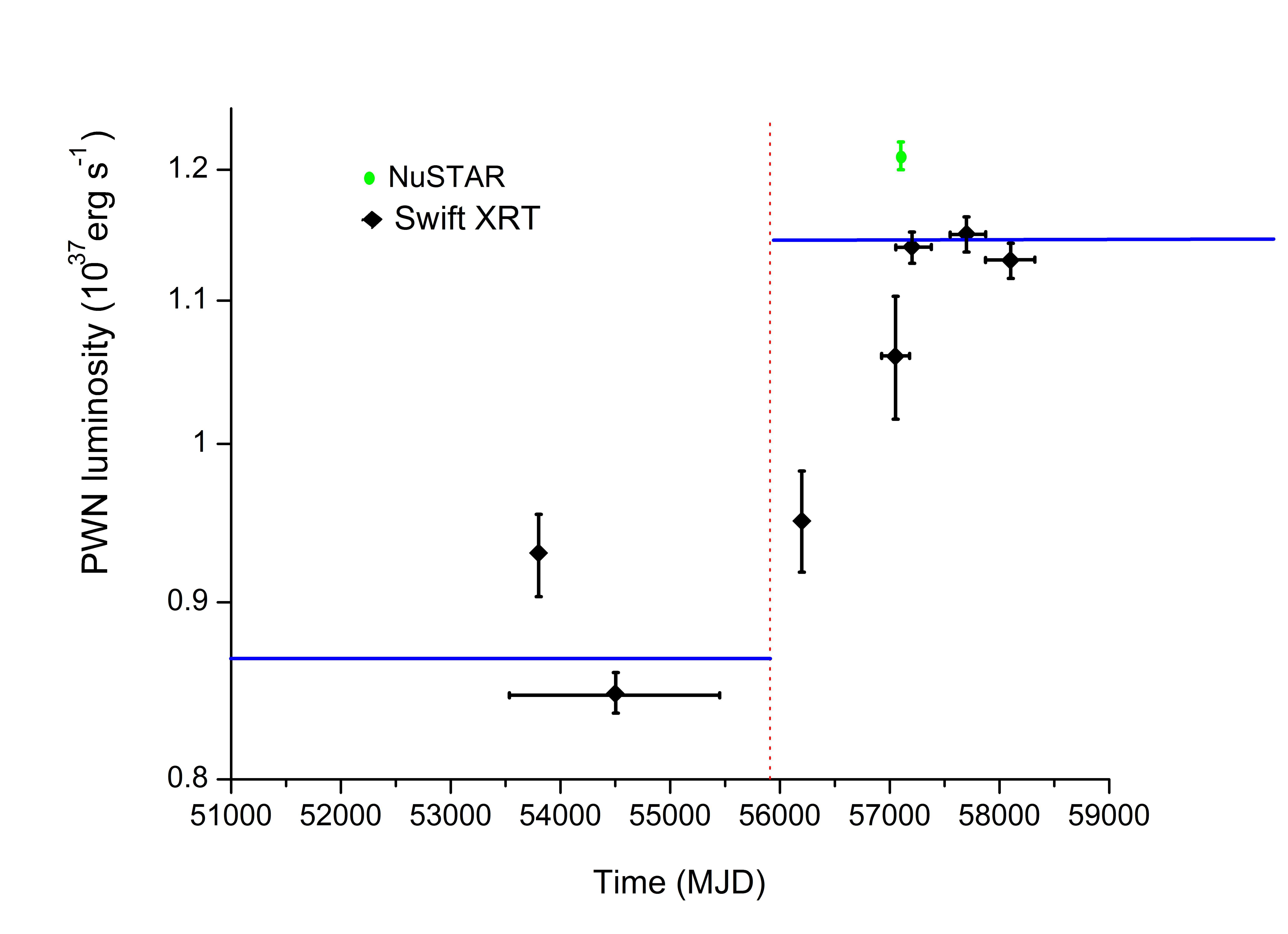}
    \caption{The jump of the X-ray luminosity detected by NuSTAR and Swift XRT related to the event of 2011. Blue lines indicate the central average values before and after the event. Note that besides its slow rise, the higher value stabilizes, arguing in favor of a permanent change in the PWN energy budget.}
    \label{fig:Lx}
\end{figure}

As a working hypothesis based in the discussion of the previous section, we will consider variations in the inclination angle $\chi$ to be proportional to the variations in the position of the ion lattice platelets in the star's surface. We can rewrite the torque equation \ref{eq:torque} in terms of the braking index:

\begin{equation}
    \frac{\dot{\chi}}{\tan{\chi}} + \frac{BR^3}{I}\left( \frac{L_p}{2c^3} \right)^{1/2} \sin^2{\chi} = \frac{(3-n)}{2}\left | \frac{\dot{\Omega}}{\Omega} \right |,
     \label{eq:brakingindex}
\end{equation}

where we consider here only a temporal variation of the inclination angle, without changes in the magnetic field total intensity or the moment of inertia. The particle luminosity before and after the SRT11 changes, but in such a short period of time that we can assume it as an instantaneous shift across the event. As can be seen in Table \ref{tab:brakingindex}, adapted from the temporal evolution of PSR B0540-69 main properties reported by \citet{wang2020braking} and \citet{rusul2023external}, before the spin-down rate transition PSR B0540-69 has an angular frequency $\nu_{\rm{bef}}=20\;\rm{Hz}$, that was unaltered across the event, but features a dramatic change after the event in the first derivative and braking index, with the latter quantity going to almost zero, and then slowly increasing in the following $\sim5$ years. \citet{wang2020braking} reported that in that same period, the X-ray luminosity of the PWN, observed by NuSTAR and Swift XRT, increased its intensity by $32\%$ (see Figure \ref{fig:Lx}), interpreted by us as evidence that the SRT11 altered the structure of the pulsar magnetosphere in such a way that the injection of energy into the PWN increased. We remind that the real emission processes in magnetospheres are highly debated \citep{melrose2021pulsar}, with a diversity of different models and assumption to understand the true spin-down mechanism(s) of pulsar. However, it is a consensus that an important feature is the electron-positron pair cascade production \citep{ruderman1975theory,timokhin2015polar}, emerging from ultra relativistic particles ($\Gamma \sim 10^7$) accelerated near the polar cap surfaces. This pair plasma flows out of the pulsar along open magnetic field lines, providing energy to the surrounding medium. Thus, we argue that an increase in the inclination angle will lead to an increase in the open lines regions and increasing the amount of energy emitted by the neutron star. This can be simply modeled if we assume that this region has a cone-like shape with a (fixed) height $h_{\rm{open}}$, which is proportional to the distance between the NS and the light cylinder region, where the magnetic field lines open. Therefore, an increase in the angle $\chi$ implies an increase in the radii $R_{\rm{open}}$ and the solid angle $\Omega_{\rm{open}}=2\pi(1-\cos\chi)$ of the open lines region, where $\chi=\arctan(R_{\rm{open}}/h_{\rm{open}})$. Moreover, assuming a simple linear relation $L_{\rm{PWN}} \approx \eta L_p$, we obtain an estimate of the variation in particle luminosity that occurred after the SRT11. This approximation is valid if the parameter $\eta$ is a function of the physical interactions between the wind particles and the nebulae matter, in turn proportional to the energy distribution of the wind, but also to the PWN density and magnetic field, both properties assumed to be  unaffected by the 2011 event due to their distance to the neutron star surface. Applying the known values before and after the SRT11 to Equation \ref{eq:brakingindex}, we determine the ratio between the initial $\chi_{\rm{bef}}$ and final $\Delta\chi+\chi_{\rm{bef}}$ angles:

\begin{equation}
    \frac{\sin(\Delta\chi+\chi_{\rm{bef}})}{\sin(\chi_{\rm{bef}})}=2.015\pm0.090.
    \label{eq:sin}
\end{equation}

Thus, any inclination angle displacement that satisfies this equation would explain the observations. An immediate corollary from this relation is that the maximum inclination angle the pulsar could have before the anomaly was $\chi_{\rm{bef}} \lesssim 30\degree$, in good agreement with the results of \citet{takata2007pulse}. Finally, to determine the conditions of the discontinuity, we need to estimate the variation $\Delta\chi$, which can be done by calculating the mechanical conditions of the crust.

Although this event observed in PSR B0540-69 is somewhat similar to a glitch, the fact that there was no change in the spin-down frequency and that the braking index did not return to the initial value are evidences that it is a related, but different phenomenon. Even so, we can approach this problem in a similar way and consider starquakes (in fact, fast-slip events, \citet{hu2023high}) as the main cause of the SRT, but with the important difference that, in this latter case, there was a simultaneous magnetic field drift. In this scenario, using 3D simulations of the magnetic field in the crust, \citet{Gourgouliatosetal2016} have shown that the local field near the surface can be much higher than the global ``dipolar" field, in fact strong enough to break and displace the lattice. Therefore, applying an analogous method of \citet{thompson1995soft} and \citet{baiko2017anisotropic}, we can consider that the breaking point is reached when the local magnetic pressure exceed the maximum elastic stress supported by the lattice and, more important, that the angle by which the magnetic field is bent $\Delta\chi\sim \delta B/B$. Therefore:
\begin{equation}
 \Delta\chi\approx \frac{4\pi}{B^2_{\rm{max}}}\mu\theta_{\rm{max}},
\end{equation}

Where $B_{\rm{max}}$ is the local field, correlated to the the maximum strain angle as shown in \citet{Horvathetal2022}: $B_{\rm{max}}\sim3\times10^{15}\left(\frac{\theta_{\rm{max}}}{0.1}\right)^{1/2}$; and the value of the shear modulus is assumed to be $\sim \, \mu \sim 10^{30}\;\rm{erg\;cm^{-3}}$, as estimated by \citet{hoffman2012mechanical} and \citet{baiko2018breaking}. Thus, the inclination angle variation is $\Delta\chi\sim 0.14\;\rm{rad}=8.00\degree$. Finally, from Equation \ref{eq:sin}, we can infer that the angle shifts from $\chi_{\rm{bef}}=(7.73\pm0.67)\degree$ to $\chi_{\rm{aft}}=(15.73\pm0.67)\degree$. At this point we must state that the actual scale of the platelets is uncertain by a numerical factor, and also that a number of platelets must be involved in the slide event. Moreover, if a group of platelets participate, the vector sum of the corresponding magnetic fields attached to each one need not be perpendicular to the surface of the crust, and a slanted magnetic field could arise after the event. As an example of this configuration, \citet{liu2025internal} modeled the torque evolution of PSR J1718–3718 and found a good fit if $\sim 142$ platelets of $\sim 30 \, \rm{m}$ each were involved. The SRT event has a degeneracy between the number of platelets and their size, at present impossible to disentangle. But the important point here is that an opening angle jumping 
by $\sim 8\degree$ is the estimate final result of the displacement $d$ of the group of platelets and the composition of the magnetic field vector in the final state.

An event related to the sudden sliding of the platelets would dissipate elastic energy, which can appear later in high-energy photons. To calculate this energy maintaining self-consistency, we can use the equations provided by \citet{thompson1995soft} to analyze the energetic of the quake phenomenon. According to the authors, the maximum magnetoelastic energy released (for a crust of volume $l^{3}$) would be converted into heat, Alfv\'en waves and seismic waves, which would dissipate on a longer timescale, depending on the viscosity of the crust:

\begin{equation}
 E_{\mathrm{max}}\sim\frac{(\delta B)^{2}}{8\pi}\times l^3=7\times 10^{36}B_{\mathrm{max}}^{-2}l_{3}^{3}\theta_{{-1}}^{2}\;\mathrm{erg}.
 \label{eq:energy}
\end{equation}

It is now known that the conversion into heat takes $\geq 90\%$ of the released energy, while both wave modes share the rest. However, to detect this emission is not guaranteed, given that the PWN energy is roughly one order of magnitude greater, and would obscure the possibility of direct observation. Moreover, the SRT was not observed in real time (the event occurred sometime between December 3rd and 17th, 2011), so any prompt emission due to the platelets cracking and fast-slip could not be immediately detected. Since the $L_{X}$ curve showed a rise followed by a stabilization of the luminosity at longer times, we interpret that the prompt release of energy 
was not the most important factor, otherwise the X-ray curve would fall again on a diffusion timescale. Thus, we argue that the shift in the angle and the reconfiguration of the magnetosphere are consistent with the observed phenomenology, since they provide a 
permanent change powering the PWN through the wind term. New observations would be important to confirm the status of the energetics and its consistence with the model.

\subsection{Possible changes in the shape of the pulse following the event}

There is in the literature firm evidence for a correlation between changes in rotational properties of pulsars (mainly $|\dot{\nu}|$) and the radio pulse shape (see \citet{lyne2010switched} and in the recent works of \citet{liu2021pulse} and \citet{liu2022pulse}, both specifically related to glitch events). We emphasize that a glitch is {\it not} what happened to Crab Twin SRT event, although the latter is a rather similar phenomenon. However, this correlated changes are not so obvious in X-ray pulsars. Only a few cases are known, like the AXP 1E2259+586 \citep{woods2004changes}, which underwent a clear change in the X-ray pulse profile during an outburst in 2002; and AXP 4U 0142+61 \citep{morii2005pulse}, where a glitch-like event followed by a significant change in the X-ray profile was detected. In the latter case, the authors suggest the possibility of a pulse profile variation caused by a crust cracking or platelet movement, that causes deformation of the crust accompanied by deformation of the magnetosphere.

If we now consider specifically the case of PSR B0540-69, a young pulsar with fast rotation and relative high magnetic field, we know that the high-frequency emission of pulsars with these characteristics is described by models where the emission region is thought to reside {\it far away} from the surface of the neutron star, e.g. at the Outer Gap (see \citet{vigano2015assessment} for a review and critique) or at the current sheet just beyond the light cylinder (\citet{lyubarsky2019radio}, \citet{petri2025double}). Considering that the work of \citet{johnston2004radio} showed radio emission from the Crab Twin in the form of giant pulses in-phase with X-ray pulses, it is plausible to suppose that both pulses are generated in the same region, and thus quite distant from the compact object. Therefore, we argue that changes in the angle between the magnetic field and the rotational axes resulting from platelet shifts only (significantly) modify the magnetosphere near the star, but may have little effect far from it. Therefore, if the pulse is generated in a distant region, it might not be possible to detect any change in the pulse structure. Inverting this argument, we may claim that the very fact that no change was observed during or after the spin-down rate transition is a proof that the pulse, whatever the mechanism of its formation, is generated far from the star, in agreement with \citet{ge2019brightening}, who used this same argument to explain the absence of changes in the pulse profile of PSR B0540-69. Our suggestion of magnetic angle change due to platelet dynamics needs a substantial angle shift, and therefore a reconfiguration of the magnetosphere to some extent, but not necessarily a change in the pulse. All these considerations would be resolved if we had a good detailed theory of how and where the pulses originate, a topic that is elusive since the very detection of the pulsar emission.\\
\\
\subsection{The braking index post-SRT11}

\begin{figure}
 \centering
 \includegraphics[width=\linewidth]{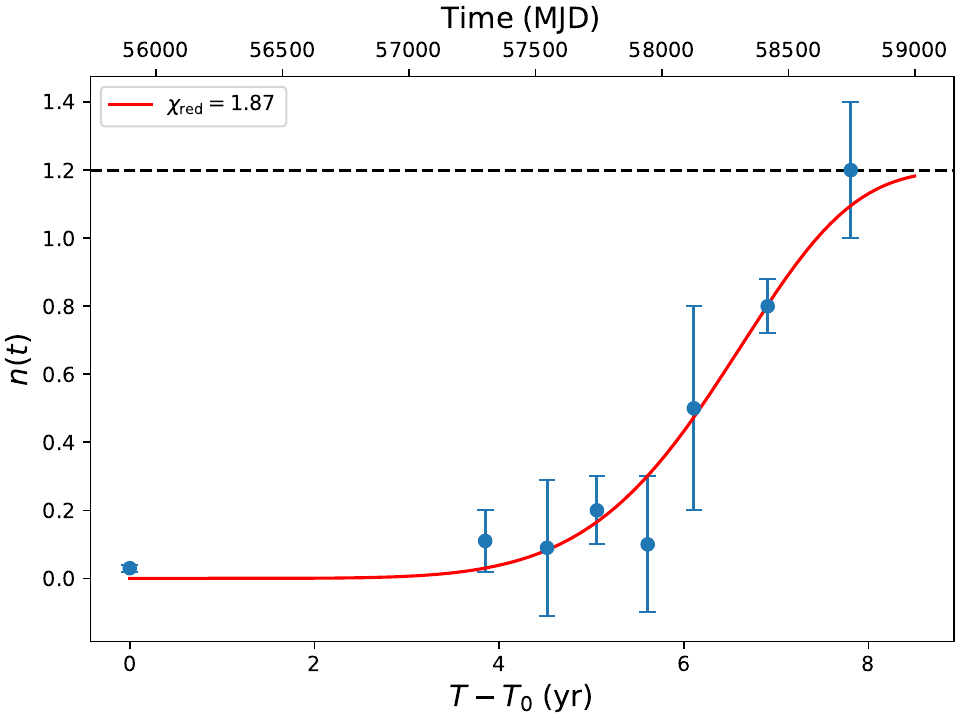}
 \caption{The best fit to the evolution of the braking index between the SRT11 time and 2019, using Equation \ref{eq:phenom}. The reduced chi-squared is $\chi_{\rm{red}}=1.87$.
 \label{fig:braking}} 
\end{figure}

Previous works related to the SRT11 event of PSR B0540-69 attempted to understand the braking index behavior between 2011 and 2019. In fact, the behavior of $n(t)$ is really peculiar and very hard to fit in most standard scenarios: after the SRT11, it remained nearly zero for five years and later, in a period of two years, abruptly increased (due to an increase of $\ddot{\nu}$, see Table \ref{tab:brakingindex}) to finally reach the constant value $n=1.2\pm 0.2$, which was maintained until 2023 (when another sudden perturbation occurred, see below) finally rising to $n=1.4$. On one hand, \citet{wang2020braking} assumed a variation in the magnetic field {\it intensity} as a trigger to the braking index disruption, and use a function with eight free parameters to describe $n(t)$; on the other hand, \citet{rusul2023external} suggest a ``pendulum-like" motion (with four free parameters) of the inner crust, resulting in an increase of the pulsar moment of inertia. We are not in a position to state with certainty which physics relates the dynamics of the platelets to the evolution of the neutron star's inclination angle, but we looked for a phenomenological function that best fits the experimental data and that has the smallest possible number of free parameters, and presume this will give us a hint of the putative processes occurring inside the neutron star. Thus, just by analyzing the braking index data, we have that $\dot{\chi}(t)$, in Equation \ref{eq:brakingindex}, must have the following properties: 1) $n(0)=3-2\left|\frac{\Omega}{\dot{\Omega}}\right|(\dot{\chi}(0)+cte.)\approx 0$, so $\dot{\chi}(0)\approx1.5-cte.$, where the constant correspond to the wind term; 2) $\dot{\chi}(t\rightarrow 8\;\mathrm{yrs})\rightarrow 0$, so from $\sim$ 2019 the braking index remains constant at $n=1.2$; and 3) the function must be concave upwards to explain the points of $n(t)$, since the latter clearly exhibits this behavior. This is the most difficult condition to satisfy, since most functions with exponential behavior, when applied in Equation \ref{eq:brakingindex}, exhibit a downwards concavity. Thus, we postulate here a stretched exponential function that is a good form to parametrize the 
change in the temporal behavior of the angle, namely

\begin{equation}
 \dot\chi(t)=Ae^{(t/\lambda)^\beta},
 \label{eq:phenom}
\end{equation}

also known as Kohlrausch–Williams–Watts (KWW) function, as the best fit to the braking index data, as can be seen in Figure \ref{fig:braking}, with the advantage of having fewer degrees of freedom than the models presented by other works. \citet{lukichev2019physical} demonstrated that The KWW function describes the motion of the overdamped linear oscillator in situations where the time ``constant" ($\lambda$ in Equation \ref{eq:phenom}) is time-dependent (it is also commonly used to describe the dielectric decay of polymers \citep{williams1970non}). Although in our problem the inclination angle displacement is related to platelet motion, and thus the KWW function is only a convenient model capable of describing the seismic dynamics (starquakes), the impact in the magnetic field evolution could offer a realistic function to model $\dot\chi(t)$. We point out that the braking index behavior indicates that after the SRT there was a long period of time (approximately 5 years) of free motion of the crust, with low viscosity/energy dissipation, tentatively associated with the slow-slip motion seen in the Earth's crust. After this period, the viscosity increased, leading to an overdamped motion, until the platelets stop and the braking index stabilize. Therefore, the KWW function must be seen as an heuristic form that satisfies the 1)-3) conditions listed above and simulate the assumed platelet motion.

\begin{table*}[t]
 \centering
 \begin{tabular}{c|cc}
       Parameter  & Inferred estimate \\ \hline
       $\chi_{\rm{bef}}$ ($\degree$) & $7.73$ \\
       $\chi_{\rm{aft}}$ ($\degree$) & $15.73$ \\
       $\Delta\chi$ ($\degree$) & $8.00$ \\
       $L_p$ ($\rm{erg\;s^{-1}}$) & $7.45\times10^{38}$\\
       $B$ ($\rm{G}$) & $4.17\times10^{13}$\\
       $\lambda$ ($\rm{s}$) &  $6.800\pm0.174$\\
       $\beta$ (-) & $6.446\pm1.626$       
 \end{tabular}
 \caption{The parameters inferred from the application of the platelet slide model and the phenomenological function \ref{eq:phenom} for the PSR B0540-69 first spin-down transition.}
 \label{tab:parameters}
\end{table*}

We now proceed by observing that, if between 2019 and 2023 the braking index remained approximately constant, then we can assume in this period $\dot{\chi}(t) \approx 0$. Thus, using the angle $\chi_{\rm{aft}}=15.73\degree$, we find $\dot{\chi}(0)\approx 6.746\times 10^{-5}\;\mathrm{rad\;yr^{-1}}$ as the initial angular displacement, which must be related to the initial velocity of the platelets. Moreover, we can determine, from Equations \ref{eq:torque} and \ref{eq:brakingindex}, both the dipolar magnetic field - $B =4.17\times10^{13}\;\rm{G}$, and the particle luminosity - $L_p=7.45 \times 10^{38}\;\rm{erg\;s^{-1}}$, the latter being an order of magnitude higher than the PWN X-ray luminosity, which suggests that only $\sim \, 10\%$ of the wind losses are converted into X-rays in the surroundings and, most importantly, that spin-down is dominated by the wind term. Even so, these values should be (by hypothesis) constant for all time after the 2011 event. Hence,  From Equation \ref{eq:phenom}, we determine $A = \left | \frac{\Omega}{\dot{\Omega}} \right |\times \dot{\chi}(0)\times \frac{1}{\tan \chi}=0.6$, as expected. Finally, using the least squares method, we also determined the best values of $\lambda$ and $\beta$, as can be seen in Table \ref{tab:parameters}.

\subsection{Other similar events}

It is relevant to mention that another SRT event (SRT23) has occurred in PSR B0540-69 in 2023 \citep{tuo2024discovery}. Again a sudden ``jump" in the braking index was reported, which went from $1.2$ to $1.4$, but with {\it decreases} in $\nu$ and $\dot\nu$. For this reason, it was called an {\it anti-glitch}. No changes in the PWN were identified after this event. If we proceed by applying the same formalism used in the SRT11, we obtain: \[\frac{\sin(\Delta \chi + \chi_{\rm{bef}})}{\sin \chi_{\rm{bef}}} = 0.92,\] which could imply in a \textit{decrease} of the inclination angle. Since this is not a consistent solution within the model of local platelet drift of \citet{Ruderman1991b}, and it is also very different from the 2011 SRT, we speculate that this could be another type of glitch-like event, perhaps without the magnetic field displacement. \citet{espinoza2024growing} examined this event and concluded that an alternative model is favored by the singular timing irregularity compared to the conventional glitch model.

Another similar anomaly was the transition of the object PSR J1846-0258 in 2006 \citep{livingstone2011post}. As in the case of PSR B0540-69, this pulsar suffered perturbations in its frequency derivatives, causing its braking index to decrease by $\sim 18\%$. An outburst was detected shortly after the corresponding SRT, which also led to an increase in the X-ray luminosity of the PWN. However, Chandra observations in 2009 indicated a brightness back to the pre-event values. Therefore, either these events are uncorrelated or the SRT did not significantly modify the structure of the magnetosphere to the point of increasing the wind luminosity, as we favor in the case of PSR B0540-69.

Finally, \citet{ge2020discovery} revealed 12 years of observations of the object PSR J1124-5916 and identified two SRTs. In the ``normal" state, the pulsar had a braking index $n=1.98\pm \,0.04$. Then, in the first SRT of 2009, the braking index also fell to values close to zero, reaching $n=0.3\pm \,0.1$, but returned to its previous value within two years. In 2012, the second low spin-down state resulted in a negative braking index $n=-2.7\pm \,0.9$, although this was considered an unreliable result due to the large timing noise and short observation time range. These examples are important for revealing the variety of state transition behaviors. Each event has peculiarities that distinguish them from other anomalies experienced by the same object or by other pulsars, and possibly only a comprehensive model that can describe the crustal dynamics and its effects on the magnetosphere of neutron stars can clarify the whole scenario.

\section{Discussion and Conclusions}

 Sudden increases in the rotational frequency of neutron stars are common in pulsars and understood within some variant of the more common glitch scenario. Several detections of this kind of phenomenon in young, rotational-powered pulsars are known (see \citet{antonopoulou2022pulsar} for a recent review). On the one hand, glitches are typically characterized by an observed spin-up of the neutron star, i.e., $\Delta\nu/\nu>0$, along with a small decrease in the spin-down rate $\Delta\dot\nu/\dot\nu\sim0.01\%$; on the other hand, it is expected that the pulsar will return to its previous state after a period of months or years. However, the peculiarities of the Crab Twin spin-down rate transition did not reveal any of these properties: the frequency was unaltered during the SRT11 event, the $\dot\nu$ step was large $36\%$ and, since 2011, PSR B0540-69 braking index never returned to the initial value $n=2.12$, thus suggesting that other factor(s) must be considered. We elaborated a simple phenomenological crustal platelet tectonics model inspired by \citet{Ruderman1991a} and the Earth's crust picture, where the neutron star accumulates a growing shear stress caused by the combination of superfluid neutron vorticity dynamics and by local magnetic pressure. When the maximum strain angle is reached, a group of platelets of the crust is expected to crack and slip along the surface, carrying with it the frozen magnetic field of each one and rearranging themselves at some distance $d$. The vector composition of the magnetic field patches and the displacement together may lead to an overall growth of the inclination angle between the magnetic field and the rotation axis. The existence of a PWN surrounding PSR B0540-69 indicates the presence of a wind, which led us to introduce an additional term related to the ejection of relativistic particles in the torque Equation \ref{eq:magneticdipole}. Hence, we have shown that an inclination angle displacement due to seismic activity in the neutron star surface, and a subsequent adjustment in the magnetosphere structure, with an increase in the amount of emitted energy by the pulsar, can simultaneously explain both the braking index anomaly post-transition and the increase in the PWN X-ray luminosity. 
 
 We also presented a phenomenological function \ref{eq:phenom} representing the underlying dynamics as the best fit to the braking index evolution in the almost 7.5 years after the sudden anomaly of 2011. Even though this is an heuristic choice, we consider that the accuracy of this fit and apparent consistency of the picture are clues for the understanding of the dynamics of a region causing the events in neutron stars. Finally, we emphasize that the spin-down rate transition event of the Crab Twin is a valuable opportunity to explore these issues related to the old idea of interior of neutron stars, platelet tectonics, magnetospheric activities and the role of winds in the general behavior of young pulsars. Starquakes, devised and elaborated to address glitches, may indeed be an appropriate model for the SRT-class events in which the rotation frequency does not change, although the torque does \citep{horvath2019braking, Sun}. The different patterns of crack-slip identified in seismologic studies of the Earth must be studied 
 thoroughly to check whether neutron star analogues occur, and which observables stem from them.

\begin{acknowledgments}
We acknowledge the support of the FAPESP Agency (S\~ao Paulo State, Brazil) through the Grant 2024/16892-2. J.E.H. thanks the CNPq Federal Agency, Brazil for a Research Scholarship over many years of activity. This study was financed in part by the Coordenação de Aperfeiçoamento de Pessoal de Nível Superior - Brasil (CAPES) - Finance Code 001.
\end{acknowledgments}

\bibliography{PSR}
\bibliographystyle{aasjournal}

\end{document}